\documentstyle[11pt,emulateapj,epsf, rotate]{article} 

\lefthead{Popov et al.}
\righthead{NEUTRON STAR CENSUS}

\begin{document}

\title {The Neutron Star Census}

\author{S.B. Popov\altaffilmark{1}, M. Colpi\altaffilmark{2}, A.
Treves\altaffilmark{3}, R.
Turolla\altaffilmark{4}, V.M. Lipunov\altaffilmark{1,5} and M.E.
Prokhorov\altaffilmark{1}}
\altaffiltext{1}{Sternberg Astronomical Institute, Universiteskii Pr. 13,
119899, Moscow, Russia; e--mail: polar@xray.sai.msu.su}
\altaffiltext{2}{Dept. of Physics, University of Milan, Via Celoria
16, 20133 Milan, Italy; e--mail: colpi@uni.mi.astro.it}
\altaffiltext{3}{Dipartimento di Scienze, Universit\`a  dell'Insubria,
Via Lucini 3,
22100, Como, Italy; e--mail: treves@uni.mi.astro.it}
\altaffiltext{4}{Dept. of Physics, University of Padova, Via Marzolo 8,
35131 Padova,
Italy; e--mail: turolla@pd.infn.it}
\altaffiltext{5}{Dept. of Physics, Moscow State University;
e--mail: lipunov@sai.msu.su}

\begin{abstract}

The paucity of  old isolated accreting neutron stars in
ROSAT observations is used to derive a lower limit on the mean
velocity of neutron stars at birth. The secular evolution of the
population is simulated following the paths of a statistical
sample of stars for different values of the initial kick velocity,
drawn from an isotropic Gaussian distribution with 
mean  velocity
$0\leq \langle V\rangle\leq 550$ ${\rm km\,s^{-1}}$. The
spin--down, induced by dipole losses and the interaction with
the ambient medium, is tracked together with the dynamical
evolution in the Galactic potential, allowing for the
determination of the fraction of stars which are, at present, in
each of the four possible stages: Ejector, Propeller, Accretor,
and Georotator. Taking  from the ROSAT All Sky Survey an upper
limit of $\sim 10$ accreting neutron stars within $\sim 140$ pc from the
Sun, we infer a lower bound for the 
mean  kick velocity, 
$ \langle V\rangle\gtrsim 200-300$ ${\rm km\,s^{-1}},$
corresponding to a velocity dispersion $\sigma_V\gtrsim
125-190$ km s$^{-1}$. The same
conclusion is reached for both a constant 
magnetic field ($B\sim 10^{12}$ G) and a magnetic field decaying exponentially with a
timescale $\sim 10^9$ yr. Such high velocities are consistent 
with those derived from radio pulsar observations. Present results,
moreover, constrain the fraction of low velocity stars, which
could have escaped pulsar statistics, to less than $1\%$.

\end{abstract}


\section{Introduction}
Isolated neutron stars (NSs) are expected to be as many as
$10^8$--$10^9$, $\sim 1\%$ of the
total stellar content of the Galaxy. 
Young NSs, active as pulsars, comprise only a tiny fraction
($\sim 10^{-3}-10^{-4}$) of the entire population, and about 1,000
have been  detected in 
radio surveys. It  is thus an observational fact
that most of the NSs remain undetected as yet.
Despite intensive searches at all wavelengths, only a few (putative)
isolated NSs which are not radio pulsars 
(or soft $\gamma$ repeaters)
have been recently
discovered in the X--rays with ROSAT (\cite{wwn96}; \cite{hetal96};
\cite{hetal98}; \cite{setal99}; \cite{moetal99}).
All these sources emit a thermal spectrum at
$\approx 100$ eV and the derived column densities place them at
relatively close distances (at most a few hundred parsecs) implying
luminosities not in excess of 
$10^{32}$ erg$\,\rm s^{-1}$. For RXJ1856-3754, 
an extremely faint optical counterpart
($m_V \sim 26$) have been firmly established by now (\cite{wm97}) while
a possible counterpart has
been found in the X--ray error box of RX J0720-3125  by  \cite{kk98} 
(see also \cite{mh98}).
For the other remaining condidates, plate searches have only placed limits
down to $m_V\gtrsim 24$. 

Although the extreme X--ray to optical flux ratio ($> 10^3$)
makes the NS option rather robust, the exact nature of their
emission is still controversial. Up to now, two main possibilities
have been suggested, either relatively young NSs
radiating away their residual internal energy or much aged NSs
accreting the interstellar gas. 
Unmagnetized, cooling models can reproduce the full spectral energy 
distribution (SED), including
the optical excess observed in the best studied object RX J1856-3754,
if  the surface temperature depends on the latitude or for Fe/Si--ash
atmospheres (\cite{wm97}; \cite{wa98}; note, however, that spectra computed
by \cite{petal96} for similar compositions fail to reproduce the observed 
SED). Accreting, pure H, atmospheres around non--magnetic NSs produce a
hard X--ray tail (\cite{ztzt95}) that is not observed in
these sources. However, as in cooling models, the hardening decreases
for increasing $B$ and the essential features of the SED 
may have a quite natural explanation in terms of magnetized, H, accretion
models (\cite{zzt99}).

Cooling NSs have short lifetimes
($\sim 10^6$ yr; \cite{page98}) and thus might be relatively rare objects.
On the other hand, at
least if the Bondi--Hoyle scenario applies, accretion can be 
severely reduced when the NS  velocity (relative to the
interstellar medium) 
exceeds  $40 $ km\,s$^{-1}$ so the process of accretion 
itself might be
unable to produce  the typical luminosities inferred from
ROSAT data.
When
available, proper motion measurements for some of the
five isolated NS candidates detected so far will prove decisive in
assessing the true nature of these sources. Meanwhile, we feel
that a more thorough analysis of the statistical properties of NSs
is of interest and can be useful in providing indirect evidence in
favor of or against the accretion scenario.

As discussed by \cite{l92},
isolated NSs can be classified into four main types:
Ejectors, Propellers, Accretors and Georotators. In Ejectors the relativistic
outflowing momentum flux is always larger than the ram pressure of the
surrounding  material so they never accrete and are
either active radio pulsars or dead pulsars, spun down by dipole
losses. In Propellers the
incoming matter can penetrate down to the Alfven radius but not
further because of the centrifugal barrier, and, although stationary
inflow can not occur, the piling up of the material at the
Alfven radius may give rise to (supposedly short) episodes of
accretion (Treves, Colpi \& Lipunov 1993 \nocite{tcl93}; \cite{pop94}).
Steady accretion is also impossible in Georotators where
(similarly to the Earth) the Alfven radius exceeds the accretion
radius, so that magnetic pressure  dominates everywhere over the
gravitational pull. It is the combination of the star period,
magnetic field, velocity and ambient medium, that decides  which type a
given
isolated NS belongs to
and, since both $P$, $B$, $V,$  and $n$ change during the star evolution,
a NS
can
go through different stages in its lifetime. This argument shows that,
at variance with what was assumed in earlier investigations (\cite{tc91};
\cite{bm93}; \cite{zetal95})
stationary accretion onto an isolated old NS (ONS hereafter) depends crucially on its
rotational
period and magnetic field.

The dynamical evolution of NSs in the Galactic potential was studied by
several
authors (\cite{pac90}; \cite{br91}; \cite{bm93}; \cite{mb94}; \cite{zetal95})
who modelled the present velocity distribution, one of the key
parameters governing accretion. Until now, however, little attention was
paid to the NSs magneto--rotational evolution. Only recently, this issue
was discussed in some detail by \cite{lxf98} and \cite{elu98},
who found that, for a given velocity
distribution, the number of accreting sources depends
strongly on the way the magnetic field decays.

Goal of this investigation is to consider these two issues
simultaneously, coupling the dynamical and the magneto--rotational
evolution for the isolated NS population. In particular, we
explore the effects induced on the current census of NSs by
varying the mean value of the kick velocity. In earlier
investigations (\cite{tc91}; \cite{bm93}; \cite{zetal95}),
the velocity distribution at birth was assumed either
Gaussian with a mean velocity modulus (inferred on the basis of existent
data, \cite{no90})
of $\sim$ 70 km$\, \rm s^{-1}$, or
skewed towards higher velocities ($\sim 200$ km s$^{-1}$) but
still rich in slow stars (\cite{pac90}).
In the last few years
however, proper motion studies and revised distance estimates of
pulsar subpopulations revealed that young neutron stars have high
mean velocities, $\sim$ 400 km s$^{-1}$ according to the original
suggestion by \cite{ll94}
or $\sim$ 200 km s$^{-1}$ on
the basis of the more recent analyses by \cite{hp97}, \cite{cc97} and
\cite{cc98}.
Although little is known about
the low--velocity tail of the distribution (\cite{h97}; \cite{hetal97}),
the possibility that the
low--velocity tail is underpopulated with respect to what was
previously assumed should be seriously taken into account. It is
our aim to revise the estimates on the number of old accreting
neutron stars in the Galaxy, and in the solar vicinity in
particular, in the light of these new data, in the attempt to
reconcile theoretical predictions with present ROSAT limits
(\cite{nt1999}).
This is of potential importance
as it may indirectly reveal how the long term evolution of key
parameters, such as the magnetic field, the period, the velocity
distribution and the star formation rate relate to the properties
of neutron stars at birth and to their interaction with the
Galactic environment.

\section{The Model}\label{model}

In this section we summarize  the main hypothesis
introduced to describe  the evolution of single stars and 
outline shortly the technique used
to explore their statistical properties, referring to \cite{pp98}
and \cite{pp99}
for details on spatial evolution calculations.

\subsection{Dynamical evolution}\label{dynev}

The dynamical evolution of each single star in the  Galactic potential is
followed
solving its equations of motion. The potential, 
proposed by \cite{mn75}
and adopted later by
Pac\'zynski (1990)\nocite{pac90}, 
is given by the superposition of a spherical halo and of
two flattened bulge/disk components. In Galactocentric cylindrical
coordinates $R$, $Z$,
$r = \sqrt{R^2 + Z^2}$, they are expressed as

\begin{eqnarray}\label{phih}
\Phi_H(R,Z)=-\frac{GM_C}{r_C}\left[\frac{1}{2}
\ln\left(1+\frac{r^2}{r_C^2}\right)+\right.\nonumber\\
\left.\frac{r_C}{r}\tan^{-1}\left(\frac{r}{r_C}
\right)\right]
\end{eqnarray}
and
\begin{equation}\label{phibd}
\Phi_i(R,Z)=-\frac{GM_i}{\sqrt{R^2+[a_i+(Z^2+b_i^2)^{1/2}]^2}}
\end{equation}
where the index $i$ stands both for $B$ (bulge) and $D$ (disk) and $H$ for halo.
The values of the parameters appearing in equations (\ref{phih}) and
(\ref{phibd}) are summarized in table \ref{table1}.

\begin{figure}
\epsfxsize=\hsize
\centerline{{\epsfbox{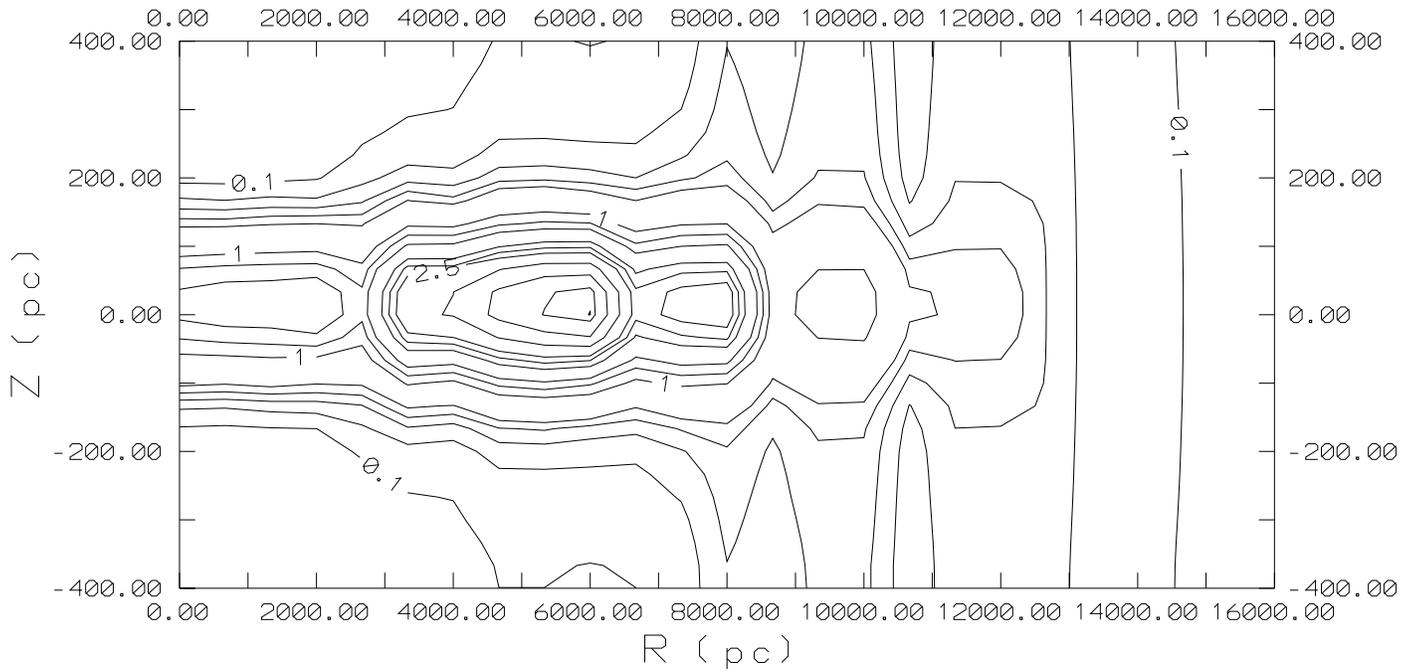}}}
\caption{The hydrogen number density distribution in the $R$--$Z$ plane.
\label{nh}}
\end{figure}

In their motion through the Galaxy, NSs interact with the
interstellar medium (ISM) when (and if) they enter the Propeller
or the Accretor phase. In these two stages the accreting material
affects significantly the star spin because braking torques arise
(see \S 2.3). Since the period evolution depends on both the star
velocity and the local density of the interstellar medium, any
attempt to investigate the statistical properties of the NS
population should incorporate a (detailed) model of the 
distribution of the ISM. Unfortunately the distribution of molecular and atomic
hydrogen in the Galaxy is highly inhomogeneous and only an
averaged description is possible. Here we use the analytical
distributions from \cite{b92}
and Zane et al. (1995) \nocite{zetal95} .
Denoting with $n_{HI}$ and $n_{H_2}$ the  neutral and molecular
hydrogen number density, the total proton density is given by
\begin{equation}
n(R, Z)= n_{HI}+2n_{H_2}.
\end{equation}
The molecular hydrogen  distribution is approximated as
\begin{equation}
n_{H_2}(R,Z)=n_2(R)\, \exp \left[ \frac{ -Z^2}{2\cdot
(70 {\rm pc})^2} \right]
\end{equation}
where $n_2(R)$ is tabulated in \cite{b92}.
The map of atomic hydrogen is more complex and for
$R \le 3.4 $ kpc we assume

\begin{equation}
   n_{HI}=n_0(R)\exp \left[ \frac{-Z^2}{2\cdot (140 \, {\rm
pc})^2}\left(\frac{R}{2\, {\rm kpc}}\right)^2 \right].
\end{equation}
The preceding expression becomes inaccurate inside $\sim 1$ kpc,
where the gas distribution is dominated by a rotating ring and
dense clouds. However, since we do not calculate the evolution of
NSs born in the central 2 kpc, this is not going to be of any
relevance (see \cite{zzt96}
for a discussion of
the diffuse X--ray emission from accreting NSs in the Galactic
Center). The density of cold and warm HI at $3.4\,  {\rm kpc} \le
R \le 8.5 \, {\rm kpc}$ can be fitted by
\begin{eqnarray}
   n_{HI}= 0.345\, \exp \left[ \frac{-Z^2}{2\cdot (212 \, {\rm pc})^2}
         \right] +\nonumber\\
          0.107\, \exp \left[ \frac{-Z^2}{2\cdot (530 \, {\rm pc})^2}
          \right] + 0.064\, \exp \left[ \frac{-Z}{403 \, {\rm pc}}\right].
\end{eqnarray}
while at radii larger than 8.5  kpc we use
\begin{equation}
   n_{HI}=n_3(R)\exp \left[ \frac{-Z^2}{2\cdot (530 \, {\rm
pc})^2}\left(\frac{R}{8.5\, {\rm kpc}}\right)^2 \right]
\end{equation}
where $n_0(R)$ and $n_3(R)$ are again taken from  \cite{b92}.
The total hydrogen density distribution used in our
computations is shown in Figure \ref{nh}; the density in the
galactic plane varies by more than one order of magnitude, ranging
from  $0.1 \, \rm {{cm^{-3}}}$ to $4.3 \,\rm {{cm^{-3}}}.$ Within
a region of $\sim 140$ pc around the Sun, the ISM is underdense,
and we take $n=0.07 \ {\rm cm}^{-3}$ (see \cite{zztt96}).

In our model we assume that the NS birthrate is constant in time
and proportional in magnitude to the square of the local gas
density (see for example \cite{c83}; \cite{ft94}).
This implies that  NSs are preferentially produced in the molecular
ring region and migrate through the Galaxy  during their
evolution, due to the large kick velocities acquired at birth.
Stars are assumed to be born in the Galactic plane ($Z=0$) and in
the range $2\, {\rm kpc} < \, R \, < 16 \, {\rm kpc}$; this
limitation has no major influence on the  results, and on the
number of sources found in the solar vicinity in particular. In
the present investigation all effects of NSs born in binary
systems (see, e.g., \cite{it98})
has been neglected. Mirroring the birthrate of massive stars, also the NS rate 
was likely to be higher in past (\cite{wg98}), so that our assumption of a 
constant NS birthrate is just a simplification. 
The effect of a time dependent NS formation rate will deserve further
study.

Neutron stars at birth have a circular velocity determined by the
Galactic potential. Superposed to this ordered motion a kick
velocity $\bf{V}$ is imparted in a random direction. The exact form of the
kick distribution is still uncertain (see, e.g., \cite{lpp96a};
\cite{it98}; \cite{bb99}).
Here we use an isotropic Gaussian distribution 
with zero mean velocity vector relative to the local
circular speed,  
and one--dimensional velocity dispersion $\sigma_V$, simply as a mean to
model
the true pulsar distribution at birth (see e.g. \cite{c98}).
The resulting distribution for the modulus $V$  is
characterized by a mean velocity 
$\langle V\rangle\equiv(8/\pi)^{1/2}\sigma_V$ 
which is varied in the interval 0--550 km\, s$^{-1}$. The velocity
dispersion of the NS progenitors, $\sim 20$ km s$^{-1}$, has been
neglected and runs with 
$\langle V\rangle$
of this order mimic a NSs population with zero 
mean kick velocity at birth.

For each star the dynamical evolution has been followed up
to present time. The evolution of the spin period $P$ (as
described in \S 2.3) is also computed, keeping memory of the
different phases the neutron star experiences (Ejector, Propeller,
etc.). A number of representative evolutionary paths have been
calculated, placing the stars at a given location in the galactic plane,
with a
randomly oriented  kick drawn from the specified Gaussian
distribution. Each evolutionary track is statistically independent, and  
gives a unique galactic orbit together with a unique history.
Each orbit is recorded every time the star crosses the boundary 
between two adjacent spatial cells in which the
Galaxy is divided. 
The time of transition between two phases is also recorded and stored. 
Approximately, a star takes  10$^6$ yr to cross a cell, which means
that the star position (and status) is recordered $\sim 10^4$ times in
a total orbital time of 10$^{10}$ yr. 
To mimic continuous star formation a single 
track is repeatedly used shifting the time ahead.
In addition,  tracks born in denser regions of the galactic plane, where
the rate of star formation
is higher,  acquires, accordingly,  a higher statistical weight. 
One orbit integration samples the behaviour of nearly
10$^4$ stars born with the same initial conditions at different times.
(we refer for details to the
``Scenario Machine'' of  \cite{lpp96b}).
A number of independent  runs have been performed using different number
of objects, up
to a maximum of $\sim 10^3$ so giving $\sim 10^7$ stellar paths.
The runs  have been carried out using different seeds
for the random orientation of kicks. Statistical fluctuations on the
number
of stars in each stage are typically of a few percent.

\subsection{Accretion physics}\label{accrphys}

The accretion rate was calculated according to the Bondi formula

\begin{equation}
\dot M=
\frac{2\pi (GM)^2 m_p n(R,Z)}{(V^2+V_s^2)^{3/2}}\simeq 10^{11}\,
n\,
v_{10}^{-3}\ {\rm g\, s}^{-1}
\end{equation}
where $m_p$ is proton mass, the sound speed $V_s$ is always 10
km$\,s^{-1}$ and $v_{10} = (V^2+V^2_s)^{1/2}$ in units of 10
km$\,
{\rm s}^{-1}$. Here and in the following $M$ and $R$ denote the NS
mass and radius, which we take equal to $1.4\, M_\odot$ and 10 km,
respectively, for all stars. Assuming, for the sake of simplicity,
a non--relativistic efficiency $\sim GM/Rc^2$, the accretion
luminosity is given by

\begin{equation}
L=\frac{GM \dot M}{R}\, .
\end{equation}
Although the spectrum emitted by accreting NSs may differ rather
substantially from a blackbody at the star effective temperature (see
\cite{ztzt95} 
for low--field spectra and \cite{zzt99}
for magnetized models) here we assume pure thermal
emission. Even in this simplified case, the effective temperature
depends on the star dipole field because accretion occurs only at the
two polar caps of radius $R_{cap}= R\sqrt{(R/R_A)}$,
\begin{equation}
R_{cap}=
9.5\times 10^3\, \mu_{30}^{-2/7}n^{1/7} v_{10}^{-3/7}\ {\rm cm}\, ;
\end{equation}
here $R_A\simeq 1.1\times 10^{10}  n^{-2/7}
v_{10}^{6/7}\mu_{30}^{4/7}\, {\rm cm}$ is the Alfven radius and
$\mu_{30}$ is the star magnetic dipole moment in units of
$10^{30}$ G cm$^{3}$ (see e.g. \cite{zztt96}; \cite{kp97}).
The reduced emitting area produces harder spectra, with an
effective temperature $\sim$ 3--4 times larger than in the
unmagnetized case
\begin{equation}
T_{eff}\simeq 5\times 10^{8} \mu_{30}^{1/7}n^{13/14}
v_{10}^{-39/14}\ {\rm K}\, ;
\end{equation}
typical temperatures are around 100 eV for $V \sim 100$ km$\, \rm
s^{-1}$
for non-evolving magnetic field, and a little bit lower for evolving
magnetic fields.

\subsection{Period evolution}\label{perevol}

All neutron stars are assumed to be born  with a period
$P(0) =$ 0.02 s, and a magnetic moment either
$\mu_{30}=1$ or $ \mu_{30}=0.5$. Different distributions
of the initial periods, like the one recently proposed by \cite{sp98},
were also tested and produced very similar results.

The ejector  regime begins with the pulsar phase and proceeds also
after the breakdown of the coherence condition when the star
becomes a dead, or silent, pulsar. In this phase the energy losses
are due to magnetic dipole radiation and the period increases in
time according to
\begin{equation}
P=P(0)+3\times 10^{-4} \mu_{30}\,\, t^{1/2}\ \rm{s}
\end{equation}
where $t$ is in yr.
When the gravitational energy density of the incoming interstellar gas
exceeds
the outward momentum flux  at the accretion radius,
$R_{ac}\simeq 2GM/v^2$, matter starts to fall in. For this condition to be
met,
the period must have reached  a critical value
\begin{equation}\label{petop}
P_{E}(E\to P)\simeq 10\, \mu_{30}^{1/2}\, n^{-1/4}\,v_{10}^{1/2}\ \rm {s}
\end{equation}
which is attained by dipole braking (for a constant field) in a time
\begin{equation}
t_{E}\simeq  10^9 \,\mu^{-1}_{30}\, n^{-1/2}\, v_{10}\ \rm{yr}\,.
\end{equation}
When $P>P_{E}(E\to P)$  matter can penetrate down to the Alfven
radius, but the interaction with the rotating magnetosphere
prevents accretion to go any further because of the centrifugal
barrier. The NS is now in the propeller phase, rotational energy
is lost to the ISM and the period keeps increasing at a rate
\begin{equation}\label{propspind}
\frac{dP}{dt}\simeq \frac{\dot M\, R^2_A \,P}{I}
\simeq K\,P^{\alpha} \ \rm{s\, s}^{-1}\, .
\end{equation}
Here we take $K=2.4\times 10^{-14} \mu_{30}^{8/7} n^{3/7}
v_{10}^{-9/7}$, $\alpha=1$ (Shakura 1975) and $I=10^{45} \ {\rm g\,
cm}^2$ is the star moment of inertia.

It should be stressed, however, that our expression for the propeller
spin--down is just an approximation. The propeller physics is very
complicated and its thorough understanding requires a full 2--D or even
3--D MHD
investigation of accretion onto a rotating dipole (see e.g. \cite{t99};
see also other spin--down formulae for that stage in 
Lipunov \& Popov 1995\nocite{lp95}).
The propeller spin--down, as modeled by equation (\ref{propspind}), is
very efficient and acts on a typical timescale
\begin{equation}
t_P\simeq 1.3\times 10^6\, \mu_{30}^{-8/7}\, n^{-3/7}\, v_{10}^{9/7}\
{\rm yr}\, .
\end{equation}
Numerical simulations (\cite{t99};
Toropin, private communication) seem
indeed to confirm that spin--down in the propeller phase is very fast
due to the large mass expulsion rate. Note that the present expression
for the propeller torque is somewhat different from that adopted by 
Treves et al. (1993; 1998)\nocite{tcl93}\nocite{tetal98}.
As the star moves through the inhomogeneous  ISM  a transition
from
the propeller back to the  ejector phase may occur
if the period attains the critical value

\begin{equation}\label{pptoe}
P_E(P\to E)\simeq 3\, \mu_{30}^{4/5} v_{10}^{6/7} n^{-2/7}
\rm{s}\, .
\end{equation}
Note that the transitions $P\to E$ and $E\to P$ are not
symmetric because the two periods (\ref{petop}) and (\ref{pptoe}) derive
from the same physical condition but evaluated at different radii, as
first discussed  by Shvartsman in the early '70s.

Accretion onto the star surface occurs when the corotation radius
$R_{co}=(GM\,P^2/4\pi^2)^{1/3}$ becomes larger than the Alfven
radius (and $R_A<R_{ac}$, see below). This implies that braking
torques have  increased the period up to
\begin{equation}
\label{paccr}
P_{A}(P\to A)\simeq 420\, \mu_{30}^{6/7}\, n^{-3/7}
\, v_{10}^{9/7}\ \rm s\, .
\end{equation}
As soon as the NS enters the accretor phase, torques produced
by stochastic angular momentum exchanges in the ISM
slow down the star
rotation to the equilibrium period
\begin{equation}
 P_{eq}=2.6\times 10^{3}\,
 v^{-2/3}_{(t)10}\,\mu_{30}^{2/3}\, n^{-2/3}\,v_{10}^{13/3}\ {\rm s}
\end{equation}
where $v_{(t)}$ the turbulent velocity of the ISM (\cite{lp95};
\cite{kp97}).

Actually the condition that $P\geq P_{A}$  is not sufficient to
guarantee that matter is captured at the accretion radius. At the
very low accretion rates expected for fast, isolated NSs, it could
be that the Alfven radius is larger than the accretion radius. The
condition $R_A<R_{ac}$ translates into a limit for the star
velocity
\begin{equation}
\label{vlim}
v <410 \, n^{1/10}\,\mu_{30}^{-1/5} \
{\rm km\, s}^{-1}\, .
\end{equation}
Above this velocity the NS behaves as a georotator and it experiences
a torque which is computed  in the same way as in the propeller stage.
The star can enter
the georotator phase only from the propeller or accretor stage.

\begin{figure}
\epsfxsize=0.9\hsize
\centerline{{\epsfbox{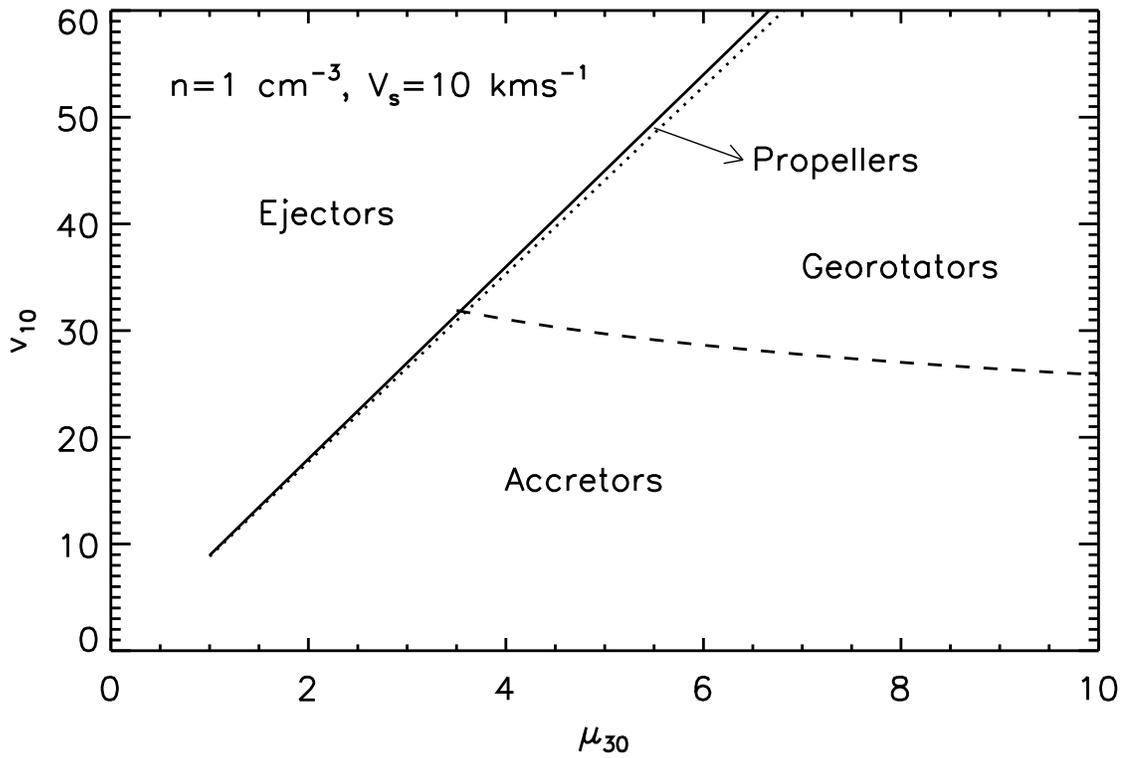}}}
\caption{The different stages of old NSs at present as a function of the
star velocity and magnetic dipole moment.}
\label{limvel}
\end{figure}

Figure \ref{limvel} illustrates the various possible stages for a NS
$10^{10}$ yr old as a function of the star velocity and magnetic field
for a constant ambient density of $1$ cm$^{-3}$. The propeller
region in Figure \ref{limvel} is very small, as a consequence of the
extremely efficient spin--down implied by equation (\ref{propspind}) with
our
choice of the parameters $\alpha$ and $K$.

\section{Results and discussion}\label{resdisc}

In this section, we present the results of our numerical simulations
both in the cases of a constant and of a decaying magnetic field.

\subsection {The NS census for a non--decaying field}

\begin{figure}
\epsfxsize=0.9\hsize
\centerline{\epsfbox{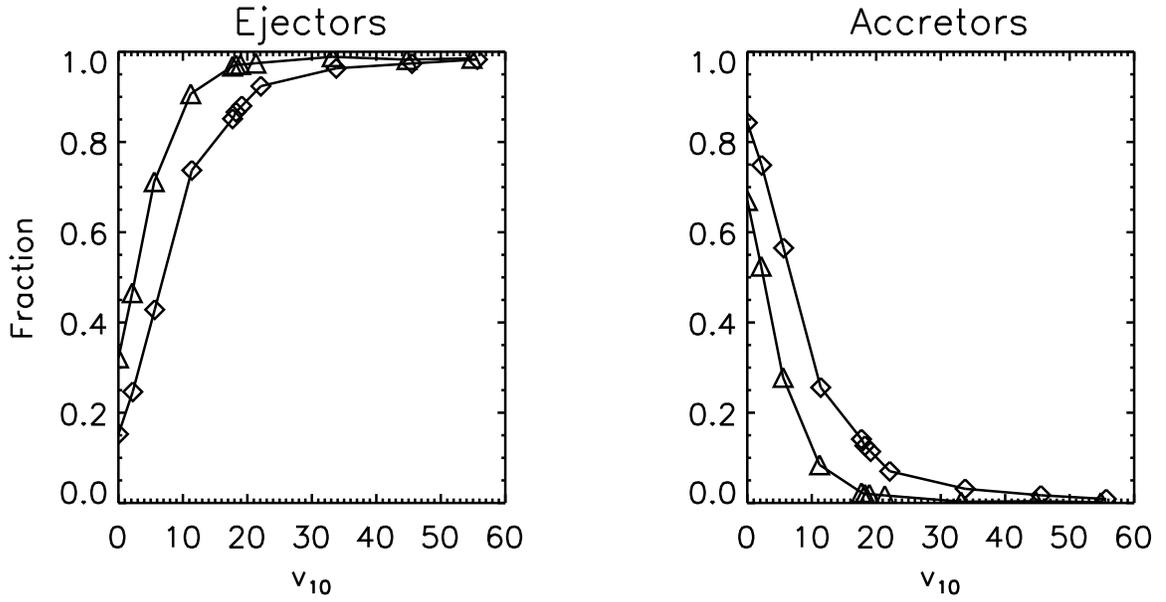}}
\caption{Fractions of NSs in the different stages vs. the mean kick
velocity for $\mu_{30}=0.5$ (triangles) and $\mu_{30}=1$
(diamonds); typical statistical uncertainty is $\sim $ 2\%.}
\label{fractions}
\end{figure}

Here we consider two values for the magnetic dipole moment
($\mu_{30}= 0.5$ and $\mu_{30}=1$) as representative of the
constant NS magnetic field. The present fraction of NSs in the
Ejector and Accretor stages as a function of the mean kick
velocity is shown  in Figure \ref{fractions}. Statistical errors
are typically $\sim 2\%$. Propellers and Georotators are much less
abundant and never exceed $\sim 1 \%$ of the total number. In
addition, their fractions  oscillate intrinsically because of the
changes in the velocity and in the ISM density along the star
trajectory. These effects add further noise to the statistical
fluctuations and, given the small number of sources in these two
states, prevent any definite conclusion about the dependence of
the fractions from the mean kick velocity.
We note that the fraction of
Accretors increases with the field strength, in agreement with the
findings of \cite{lxf98} and \cite{elu98}. 
Highly
magnetized NSs suffer, in fact, a more severe spin--down during
the Ejector and Propeller phases and reach the accretion phase
earlier in their history. For large enough mean kick velocities
($> 400$ km s$^{-1}$ for $\mu_{30}=1$ and $> 300$ km s$^{-1}$ for
$\mu_{30}=0.5$) the fraction of accretors becomes very small and
again results become statistically uncertain. We note also that
the numerical simulation produces a picture of the present status
of the Galactic NS population which is consistent with that
emerging from Figure \ref{limvel}. In fact, if we assume that 
the velocity gives a measure of the 
average kick, we see, with
reference to $\mu_{30}=1$,  that Accretors are more abundant
than Ejectors up to $\sim 100 \ {\rm km\, s}^{-1}$, Propellers are
extremely rare and Georotators nearly absent. This is precisely
what Figure \ref{fractions} shows.

\begin{figure}
\epsfxsize=0.9\hsize
\centerline{\epsfbox{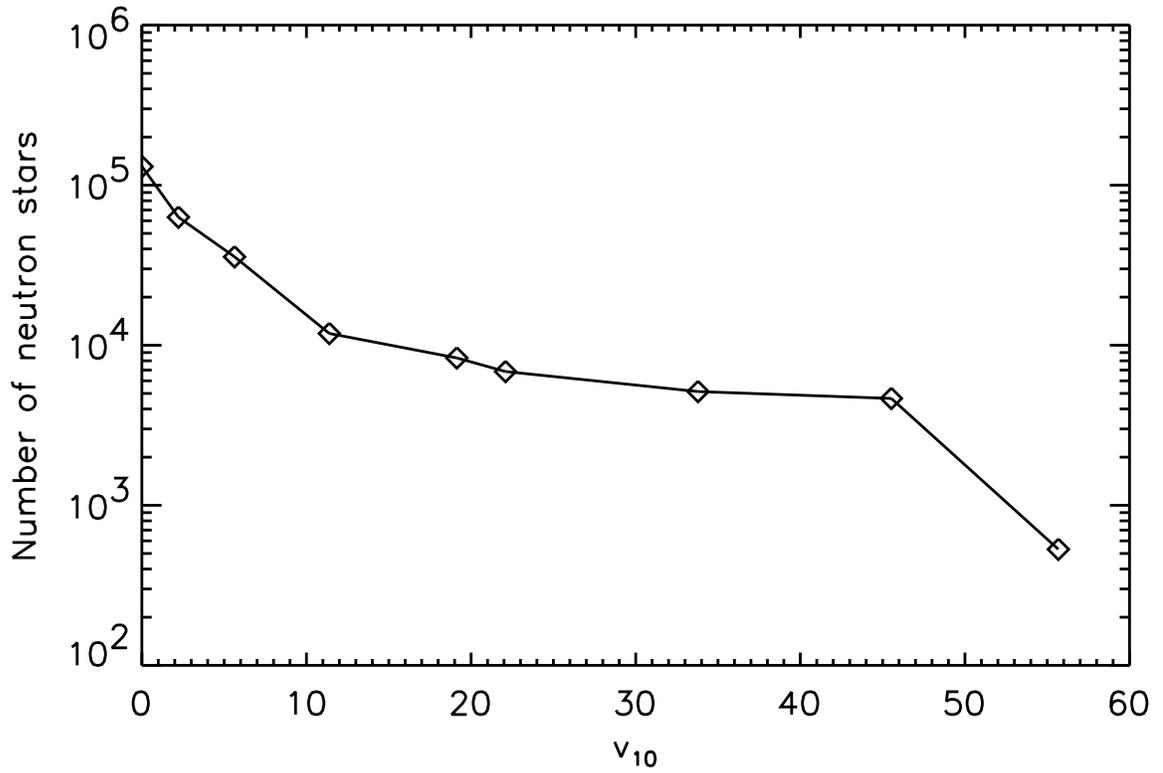}}
\caption{Total number of NSs in the Solar vicinity ($R<140$ pc) for
$\mu_{30}=1$.}
\label{nsuntot}
\end{figure}

To obtain an estimate of the number of accreting sources in the
Solar vicinity (taken to be a sphere of radius 140 pc centered on
the Sun), we used all the stars contained in a torus of the same
radius located at 8 kpc from the center of the Galaxy. The result
was then rescaled to the volume of interest and is shown in Figure
\ref{nsuntot}. Here, and in the following the total number of
Galactic NSs was assumed to be $10^9$. As expected, the local
density is a sensitive function of the kick velocity and varies
between $n\sim 8.5\times 10^{-3}$ pc$^{-3}$ and $n\sim 6\times
10^{-4}$ pc$^{-3}$ for 
$\langle V\rangle= 0$ and 190 km\,
${\rm s}^{-1}$ respectively. These figures should be compared with
$n\sim 1.4\times 10^{-3}\, (N/10^9)$ pc$^{-3}$ as derived by 
Paczy\'nski (1990) \nocite{pac90}
for a zero mean velocity and with  $n\sim
3-7\times 10^{-4}\, (N/10^9)$ found by Blaes \& Madau (1993) 
\nocite{bm93} and Zane et al. (1995) \nocite{zetal95} 
for an initial mean velocity of $\sim 60$
km s$^{-1}$. The values of the local density of isolated NSs are
thus consistent with these previous estimates.

The density of isolated NSs projected onto the Galactic plane is
$\sim 2.4\times 10^5\, (N/10^9)$ kpc$^{-2}$ for average 
initial velocities $\sim
200$ km s$^{-1}$ and a scaleheight of 200 pc. This figure is
close to the one deduced by  Ne\"uhauser \& Tr\"umper (1999) \nocite{nt1999}
from radio pulsars observations (Lyne et al. 1998) with the assumption of a
pulsar lifetime $\sim 10^7$ yr.
We would like to note, however, that a projected density $\sim 10^5$
kpc$^{-2}$ is now obtained for $N=10^9$ instead of
$10^8$ (as in \cite{mb94};
see also \cite{nt1999})
because of the larger mean velocity at birth. A
total number $\sim 10^9$ appears to be consistent with the nucleosynthesis
and chemical evolution of the Galaxy, while $10^8$ is derived
from radio pulsars observations. For the time being, it is still
uncertain if all NSs experience an active radio pulsar phase, due to
low or unusually high initial magnetic fields or/and long periods
(\cite{Gott98}), or to 
the fall--back in the aftermath of the supernova explosion (\cite{fall96}; 
\cite{GPZ99}). Radio pulsars are observed only in a fraction of SNRs (see
e.g. \cite{k96}; \cite{f98}),
and even some of these coincidences are doubtful, so
there is a serious possibility that the total number of NSs derived
from radio pulsar statistics is only a lower limit.

\begin{figure}
\epsfxsize=0.9\hsize
\centerline{\epsfbox{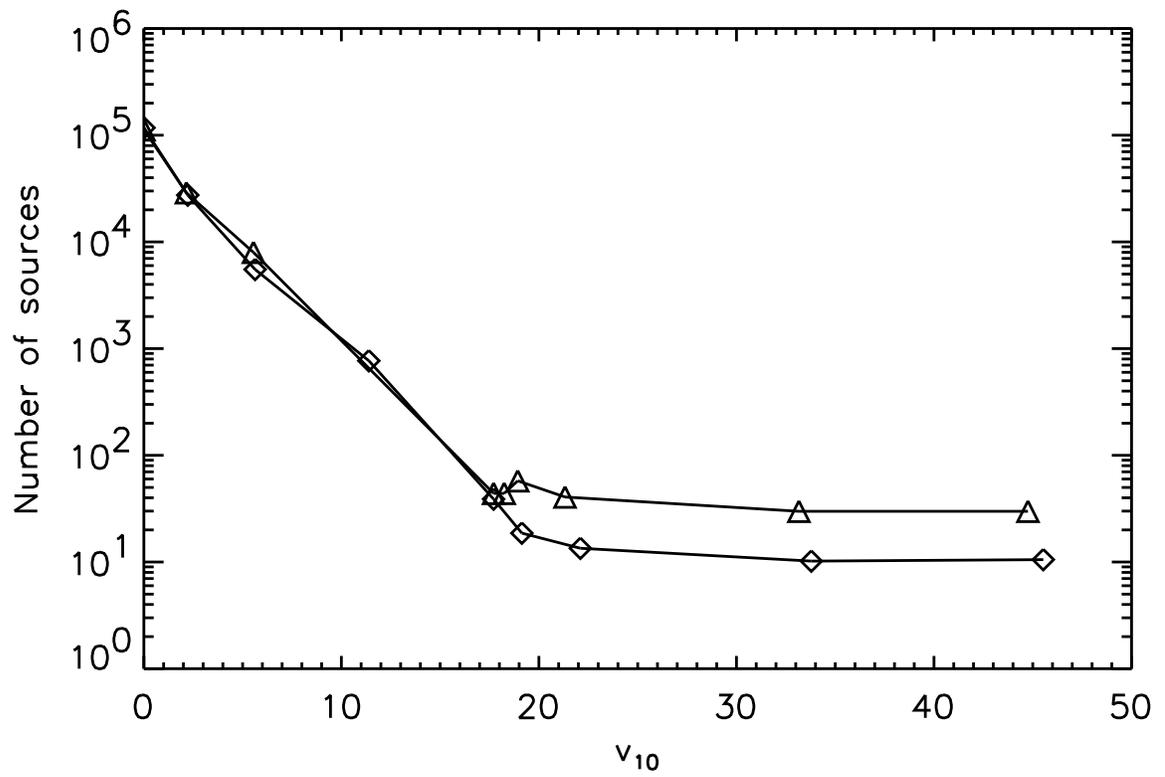}}
\caption{Number of accreting NSs in the Solar vicinity above the ROSAT
All Sky Survey
detection limit for a constant ($\mu_{30}=1$, diamonds) and decaying field
($t_d = 2.2\times 10^9 $ yrs, triangles).}
\label{nsunobs}
\end{figure}

In order to compare the expected number of accreting ONSs with the
ROSAT All Sky Survey (RASS) results, we evaluated the number of
those ONSs, within 140 pc from the Sun, producing an unabsorbed
flux of $10^{-13}$ erg cm$^{-2}$ s$^{-1}$ at energies $\sim 100$
eV (the contribution from sources at larger distances is found to
be negligible). This flux limit should correspond to a RASS count
rate of 0.01 cts s$^{-1}$ for a blackbody spectrum, once the
interstellar absorption and the ROSAT response function are
accounted for. The results are illustrated in Figure \ref{nsunobs}
(above  200 km s$^{-1}$ statistical errors are dominant). As
expected, the dependence of the number of visible sources on the
kick velocity is rather strong. The main result is that for mean
velocities below 200 km s$^{-1}$ the number of  ONSs with a flux
above the RASS detection limit would exceed 10. Most recent
analyses on the number of isolated NSs in the RASS (\cite{nt1999})
indicate that the upper limit is below 10. This
implies a {\it lower limit} for the  
mean kick velocity at birth of
$\sim 200$ km s$^{-1}$ for a total number of stars $\sim 10^9$.

This result is in  agreement with the estimates derived from
pulsar statistics. An important aspect is that our results exclude
the possible presence of a low--velocity tailat birth in  excees to that
contained in a Gaussian with 
$\langle V\rangle \sim 200$ km s$^{-1},$ i.e., the 0.05$\%$
($=(2\pi)^{-(3/2)}[V/\sigma]^3$)
 of the whole
population, when considering  stars having  $V<$  40 km
s$^{-1}.$

\subsection {The NS census for a decaying field}

The time evolution of the magnetic field in isolated NSs is still
a very controversial issue and no firm conclusion has been
established as yet (\cite{bhatta92}). A strong point is that radio pulsar
observations (see e.g. \cite{ll98})
seem to rule out fast
decay with typical times less than $\approx 10$ Myr, but this does
not exclude the possibility that $B$ decays over much longer
timescales ($t_d \sim 10^9-10^{10}$ yr). For this reason we have
investigated to what extent the decay of the $B$--field influences
the results presented in the previous section. We refer here only
to a very simplified picture in which $B(t) = B(0)\exp{(-t/t_d)}$
and no attempt is made to justify this law on a physical ground.
(See \cite{ppB99} for a more complete study in the parameter space.)
In this respect we just mention that detailed models predict both
exponential  (quite similar to the one assumed here) and
non--exponential decay (e.g. a power--law, \cite{uk97}). 
Calculations have been performed for $t_d=1.1\times 10^9$
yr, $t_d=2.2\times 10^9$ yr and $\mu_{30}(0) =1$. Since no bottom
field was specified, the magnetic moment becomes very low for
stars born $\sim 10^{10}$ yr ago ($\sim 10^{26}$ G\, cm$^3$ and
$\sim 10^{28}$ G\,cm$^3$ for the two values of $t_d$ respectively).

\begin{figure}
\epsfxsize=0.9\hsize
\centerline{\epsfbox{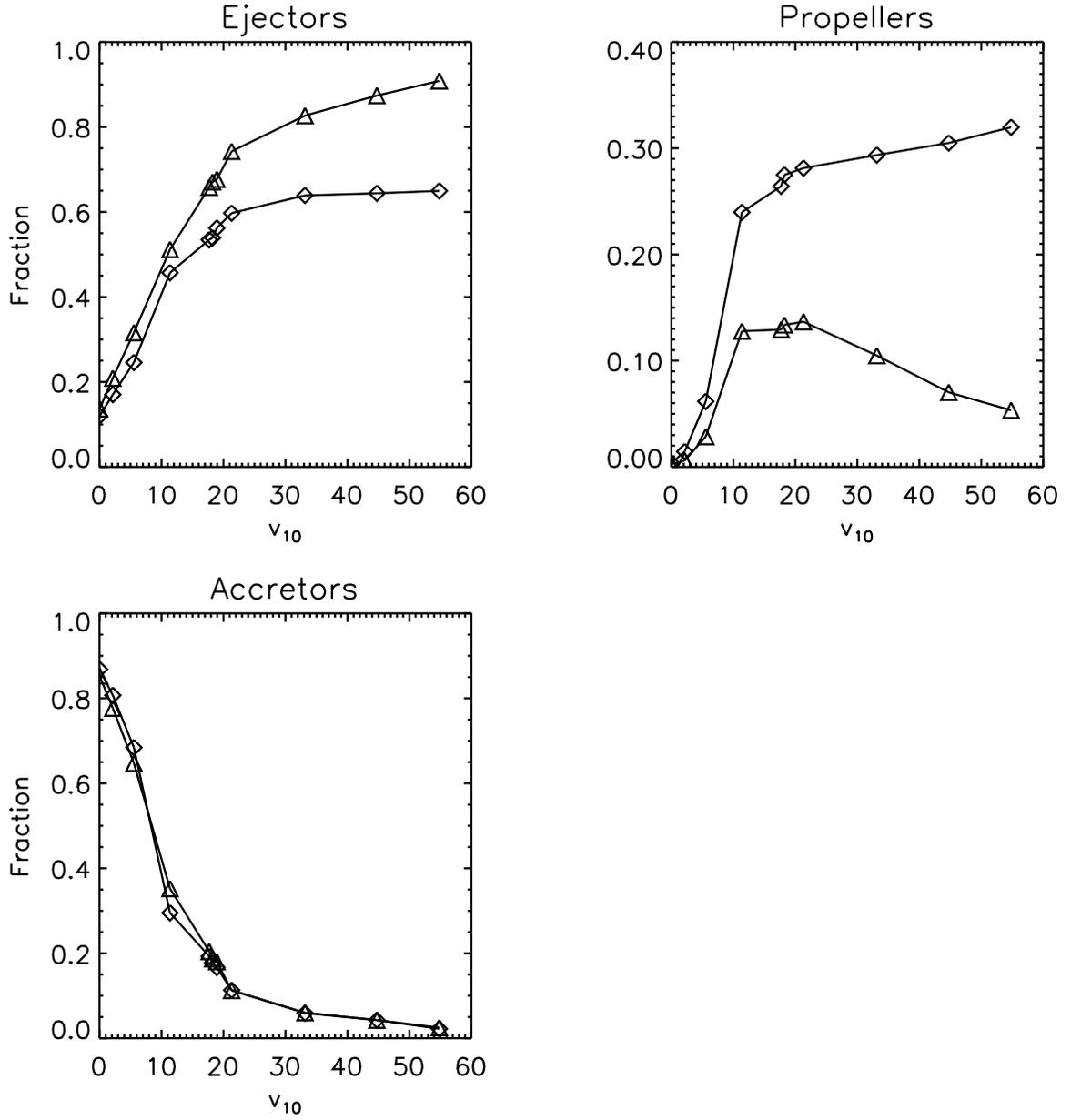}}
\caption{Fractions of NSs in the different stages vs. the average kick
velocity for a decaying field with an e--folding time
$t_d=2.2\times 10^9$ yrs (triangles) and $t_d=1.1\times 10^9$ yrs
(diamonds).}
\label{fractions_decay}
\end{figure}

Results are summarized in Figure \ref{fractions_decay}. As it is
expected (see Colpi et al. 1998\nocite{elu98}), the
number of Propellers is significantly increased with respect to
the non--decaying case, while Ejectors are now less abundant.
Georotators are still very rare, $\lesssim 1 \%$, and are not
shown in Figure \ref{fractions_decay}. The fraction  of Accretors
is approximately the same for the two values of $t_d$, and, at
least for low mean velocities, is comparable to that of the
non--decaying field while,
at larger  speeds, it seems to be somehow higher.
This shows that the fraction of Accretors 
depends to some extent on how the magnetic field decays.  
We know  that a core field decaying over a time $\sim 10^8$ yr before   
freezing would produce an underabundance of Accretors relative to
the case of a constant  field (Colpi et al. 1998\nocite{elu98}) because
NSs persist in
the Ejector or Propeller phase never approaching  the  Accretor phase.
By contrast, a  fast and progressive decay of $B$ 
would lead to an overabundance of Accretors because this situation is similar to
``turning off'' the magnetic field, i.e., quenching any magnetospheric
effect
on the infalling matter. 

Summarizing, we can conclude that, although both the initial distribution
and the subsequent evolution of the
magnetic field strongly influences the NS census altering 
the fraction of Ejectors relative to Propellers, the lower bound on the 
average kick derived from ROSAT surveys is not
very sensitive to $B$, at least for not too extreme values of $t_d$ and
$\mu(0)$, within this model.

\section{Conclusions}

In this paper we have investigated how the present distribution of
neutron stars in the different stages (Ejector, Propeller,
Accretor and Georotator) depends on the star mean velocity at
birth. On the basis of a total of $\sim 10^9$ NSs, the fraction of
Accretors  was used to estimate the number of sources within 140
pc from the Sun which should have been detected by ROSAT. Most
recent analysis of ROSAT data indicate that no more than $\sim 10$
non--optically identified sources can be accreting ONSs. This
implies that the 
average velocity of the NS population at birth has to
exceed $\sim 200 \ {\rm km\, s^{-1}}$, a Figure which is
consistent with those derived from radio pulsars statistics. We
have found that this lower limit on the mean kick velocity is
substantially the same either for a constant or a decaying
$B$--field, unless the decay timescale is shorter than $\sim 10^9$
yr. Since observable accretion--powered ONSs are slow objects, our
results exclude also the possibility that the present velocity
distribution of NSs is richer in low--velocity objects with
respect to a Maxwellian. The paucity of accreting ONSs seem
therefore to lend further support in favor of neutron stars as
being very fast objects spending most of their live in the Ejector phase.

\section*{Acknowledgments}
We are thankful to the referee, Dr. F.M. Walter
for his comments and a critical reading of the manuscript.
Work partially supported by the European
Commission under contract ERBFMRX-CT98-0195.
The work of S.P., V.L and M.P. was supported by grants
RFBR 98-02-16801 and INTAS 96-0315.
S.P. and V.L. gratefully acknowledge the University of Milan and
of Insubria (Como) for support during their visits. 

\clearpage


\begin{deluxetable}{ccc}
\tablecolumns{3}
\tablewidth{0pt}
\tablecaption{Parameters for the various contributions to the Galactic
potential\label{table1}}
\tablenum{1}
\tablehead{
\colhead{Halo} &
\colhead{Bulge} &
\colhead{Disk}\nl }
\startdata
 $r_C = 6.0$ kpc & $a_B=0.0$ kpc & $a_D = 3.7$ kpc \\
 -- & $b_B = 277$ pc & $b_D = 200$ pc \\
 $M_C = 5.0\times 10^{10} \, M_\odot$ & $M_B = 1.1\times 10^{10} \,
M_\odot$ & $M_D = 8.1\times 10^{10} \, M_\odot$ \\
\enddata
\end{deluxetable}

\clearpage

\end{document}